\documentclass[aps,twocolumn,float,prd,psfig,amsmath]{revtex4}
\usepackage[dvipdfmx]{graphicx}
\usepackage{amsmath,amssymb,natbib,siunitx,booktabs,url}
\usepackage{color}

\newcommand{\add}[1]{\textcolor{black}{#1}}

\newcommand{\lsim}{\lesssim}

\newcommand{\vect}[1]{\mbox{\boldmath${#1}$}}
\newcommand{\lmk}{\left(}
\newcommand{\rmk}{\right)}
\newcommand{\lnk}{\left\{ }
\newcommand{\rnk}{\right\} }
\newcommand{\lkk}{\left[}
\newcommand{\rkk}{\right]}
\newcommand{\lla}{\left\langle}

\newcommand{\rra}{\right\rangle}

\newcommand{\beq}{\begin{equation}}
\newcommand{\beqa}{\begin{eqnarray}}
		  \newcommand{\eeq}{\end{equation}}
\newcommand{\eeqa}{\end{eqnarray}}

\begin{document}

\title{Exploring the Orbital  Alignments of Galactic Close White Dwarf Binaries with LISA}

\author{Naoki Seto }
\affiliation{Department of Physics, Kyoto University, 
Kyoto 606-8502, Japan
}

\date{\today}

\begin{abstract}
Using the proposed space gravitational wave detector LISA, we will be able to measure the geometrical configurations of $\sim 10^4$ close white dwarf binaries in our Galaxy.  The obtained data will be an  entirely new resource to    examine the randomness of their orbital orientations.  Partly motivated by a recent reported on the systematic alignments  of  bulge planetary nebulae,  we discuss the outlook of the  orientational analysis with LISA.     We  find that a quadrupole  pattern as small as $\sim 0.05$  can be detected for bulge white dwarf binaries, owing to their large available number.  
From such a pattern analysis, we might geometrically explore fossil records in our Galaxy billions of years ago. 
 
\end{abstract}

\maketitle

\section{introduction}
Since 2015, the LIGO-Virgo-Kagra network has detected gravitational wave (GW) signals from $\sim 100$ merging extra-Galactic binaries in the 10-1000Hz band \cite{LIGOScientific:2016aoc,LIGOScientific:2017vwq,KAGRA:2021vkt}.
Quite recently, in the nHz band,   various theoretical models have been actively  discussed  with the advent of new  pulsar timing data \cite{NANOGrav:2023gor,EPTA:2023fyk,Reardon:2023gzh,Xu:2023wog}.  
  In the 2030s, the Laser Interferometer Space Antenna (LISA) will be  launched and will explore GWs around 0.1-100mHz \cite{LISA:2022yao}.  LISA has the potential to observe massive black holes   at cosmological distances, although the estimated detection rates have large uncertainties \cite{LISA:2022yao}.

More securely, LISA will separately detect $\sim 10^4$ close white dwarf binaries (CWDBs) in our Galaxy, as nearly monochromatic GW sources \cite{LISA:2022yao,Hils:1990vc,Nissanke:2012eh,Seto:2022iuf,Kupfer:2018jee}.  Indeed, for improving the effective sensitivity of LISA,  it is essential to identify these  vast number of binaries and subtract their foreground  GW signals \cite{Littenberg:2020bxy}. 
A significant fraction ($\sim 30$\%) of the  identified CWDBs will be the bulge component, located near the Galactic center \cite{Ruiter:2007xx}.  The remaining ones will be  distributed more broadly around the Galactic disk. 
In both cases, the orbital motions of the CWDBs are directly  encoded in the emitted GW signals \cite{pw,Cutler:1997ta,Takahashi:2002ky}.  Using  LISA, we can   receive the GW signals, decode them and generate a long and  high-quality list for the orbital  configurations of the  CWDBs, solely based on the first principles of physics.   Note that most of CWDBs are expected to have  negligible eccentricities due to tidal effects (as for the known CWDBs \cite{LISA:2022yao,Kupfer:2018jee}).

In relation to the orientations (i.e.  the directions of angular momentum vectors) of CWDBs,    on the basis of electromagnetic observations, there recently appeared an interesting report  on  the bulge planetary nebulae (PNe) which  are (or are inferred to be) specifically associated with short-period ($\lsim $1\,day) binaries  (in total of 14 systems) \cite{Tan}. In contrast to the whole bulge PN populations, the orientations of the symmetric axes of the specific PNe  show concentrations nearly parallel to the Galactic plane,  at 5$\sigma$ significance. PNe are ionized gas ejected at the formation of white dwarfs \cite{kwi} and closely related to the common envelope phases \cite{Ivanova:2012vx}.  In the report, the orientations of the specific subset are considered to be parallel to the orbital angular momentum vectors  of the associated short-period binaries (see also \cite{hil}). Then, the angular momentum vectors of the binaries are not randomly oriented but more probable to be nearly parallel to the Galactic disk  (roughly speaking, removing PNe from the triple geometrical relations between  the short-period binaries, PNe and the Galactic plane). 
In general, binary formation should involve various physical processes and could also depend on time and location \cite{Duchene:2013cba}.  The reported alignments  might be embedded already at the formations of the bulge binaries billions of years ago, e.g. due to  ordered strong magnetic fields as argued in the report \cite{Tan}.

Regarding the potential alignments of compact binaries in the bulge, we await further studies, in particular, independent observational  analyses  (see also \cite{huang} for  nearby binaries).  Here the long list of CWDBs  provided by LISA   could be an invaluable and solid resource.
With the large number of the available sample, we will be able to detect a weak anisotropic pattern existing  in their orientation distribution function.  

In this work, 
we discuss the outlook of such an inquiry with LISA, paying special attentions  to the geometrical aspects of the involved systems. Along the way, we point out a fourfold degeneracy at determining  the polarization angle $\psi$, which fixes the binary's orientation around the line-of-sight direction.   This degeneracy is  induced by the underlying symmetry of the measurable gravitational waveform and can be effectively regarded as an irreversibility in the  information transfer (from the encoding to the decoding),  partially hampering  our observational analysis.  
  For the data analysis of  actual CWDB sample,  we examine  a simple dualistic approach to statistically enhance its anisotropic  pattern.    The associated detection limit can be as small as $\sim 0.05$ for a quadupole mode of the spherical harmonic expansion.   Through this pattern analysis, 
LISA  might enable us to geometrically delve into the ancient history of our Galaxy.

\section{orientations of CWDBs }
\subsection{Axisymmetric Model}
As shown in Fig. 1, 
we first  define the relevant unit vectors for describing  the configuration of  a circular CWDB.  We put  its sky direction $\vec n$ and its orientation $\vec j$  (parallel to its angular momentum vector).  We also set $\vec q$ as the direction of the Galactic rotation axis.  The Galactic plane is normal to $\vec q$.

Given the recent report on the bulge PNe \cite{Tan}, we are primarily interested in the probability distribution function ${\cal P}(\vec j)$ for the orientations $\vec j$ of the bulge CWDBs,   in particular, its  pattern   relative to the rotation axis  $\vec q$.   
  We thus set $\vec q$ as the polar direction for the spherical harmonic bases $Y_{lm}(\vec j)$.

We  assume that the function ${\cal P}(\vec j)$ is axisymmetric around  the vector  $\vec q$.   Then the  function ${\cal P}(\vec j)$ depends only on the polar angle $\theta$  (i.e. only with $m=0$ modes).  Given the normalization condition,  the resultant axisymmetric   function is expanded as 
\beqa
{\cal P}_A(\cos \theta )=&&(4\pi)^{-1/2} [Y_{00}(\theta)+a_{10} Y_{10}(\theta)+a_{20} Y_{20}(\theta)\nonumber\\
 & & +a_{30} Y_{30}(\theta)+a_{40} Y_{40}(\theta)+\cdots] \label{pdf0}
\eeqa
defined in the range 
$0\le \theta\le \pi$.  We have $Y_{l0}(\theta)\propto L_l(\cos\theta)$  with  the Legendre polynomials $L_l(x)$ (in an unconventional notation  to prevent  confuses with probability distribution functions), \add{ which satisfy the odd-even identities 
\beq
L_l(-x)=(-1)^l  L_l(x). \label{eq2}
\eeq}
  We present some of the  explicit forms 
$Y_{00}=(4\pi)^{-1/2}$, $Y_{10}\propto \cos\theta$,  $Y_{20}\propto (3\cos^2\theta -1)/2$, $Y_{30}\propto (5\cos^3\theta -3\cos)/2$ and  $Y_{40}\propto (35\cos^4\theta -30\cos^2\theta+3)/8$.  For a  even $l$, we have  $Y_{l0}(0)=Y_{l0}(\pi)>Y_{l0}(\pi/2)$, and  
a negative value $a_{l0}$ resultantly induces a higher concentration to the equatorial directions $(\theta=\pi/2)$ rather than the polar directions. We have ${\cal P}(\vec j)=1/(4\pi)$ for the isotropic (random) orientation distribution with $a_{10}=a_{20}=\cdots=0$.  \add{Later, in Sec. III, we will discuss which parameters $a_{l0}$ we can determine for the Galactic CWDBs with LISA. }

\begin{figure}
 \includegraphics[width=0.8\linewidth]{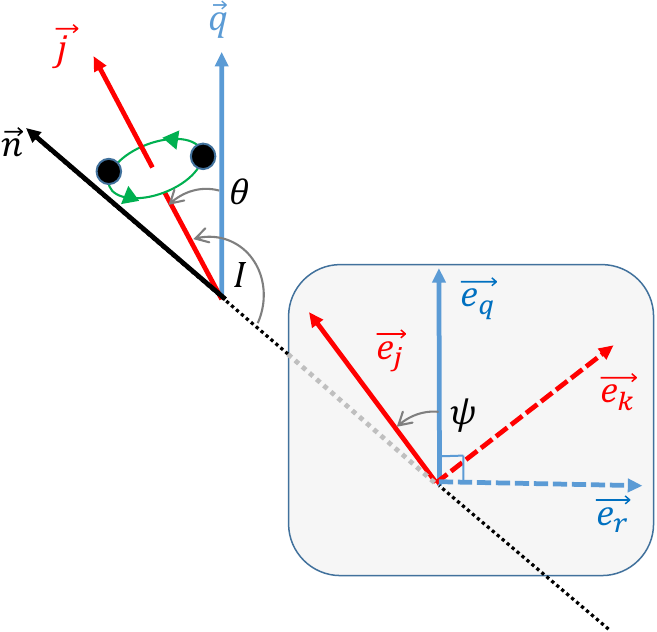} 
 \caption{Schematic picture for a binary configuration. All the seven vectors are unit vectors.  The vectors  $\vec n$  and $\vec j$ represent the direction and  orientation of the binary.  The Galactic plane is normal to the vector $\vec q$.  The angles $I$ and $\theta$ are respectively  between  $\vec j$-$\vec n$ and $\vec j$-$\vec q$.
   The gray plane is normal to the line-of-sight direction  $\vec n$, and  the vectors $\vec e_j$ and $\vec e_q$ are the projections of  $\vec j$ and $\vec q$, with the polarization angle $\psi$ between them.  The two remaining vectors $\vec e_k$ and $\vec e_r$ are respectively perpendicular to $\vec e_j$ and $\vec e_q$.    
 }  \label{fig:volume}
\end{figure}

\subsection{Coordinate Transformation}

For discussing GW observation, it is convenient to introduce the two unit  vectors  $({\vec e_j},{\vec e_k})$ normal to the binary direction $\vec n$  (see Fig. 1).   Here   the unit vector  $\vec e_j$ shows the transverse projection of the orientation vector $\vec j$,  with the remaining one $\vec e_k(={\vec n} \times{\vec e_j})$.  

 The unit vector $\vec e_q$ is similarly defined by the transverse projection of $\vec q$ with  $\vec e_r(={\vec n} \times{\vec e_q})$. For a given binary direction $\vec n$, the vectors $\vec e_q$ and $\vec e_r$ are fixed, and we can specify the orientation $\vec j$ in terms of the inclination angle $I$ and the polarization angle $\psi $ (see Fig.  1).   
We have the following relations
\beq
{\vec e}_k={\vec e}_r\cos\psi+{\vec e}_q \sin\psi,~
{\vec e}_j=-{\vec e}_r\sin\psi+{\vec e}_q \cos\psi  .  \label{be1}
\eeq

Our next task is to generate the distribution function $p(I,\psi)$ given in the observer's frame,  using  the  aforementioned one  ${\cal P}_A(\cos \theta )$ given in the Galactic frame.
To this end, we first discuss the mismatch between $\vec q$ and $\vec e_q$. Our solar system is almost on the mid Galactic plane.  The distance to the bulge is $\sim 8.3$kpc and its scale height is $\sim 0.5$kpc \cite{ga}.  Therefore, the  vector $\vec q$ is nearly normal to $\vec n$, and  we have $\vec q\simeq \vec e_q$ with  the typical mismatch angle (in radians)  $\gamma\sim 0.5/8.3\,{\rm }\ll 1$.   We then have 
\beq
\cos \theta={\vec j}\cdot {\vec q} \simeq {\vec j}\cdot {\vec e}_q =\sin I \cos\psi \label{tr1}
\eeq
and correspondingly 
\beq
p(I,\psi)\simeq{\cal P}_A (\sin I \cos\psi ). \label{tr2}
\eeq
Below, we apply  equalities to the relations (\ref{tr1}) and (\ref{tr2}) (a distant observer approximation). 
It is a straightforward but cumbersome task to derive the function  $p(I,\psi)$ without the approximation. 
 Importantly, our approximation does not introduce artificial anisotropies to $p(I,\psi)$ from the originally isotropic function ${\cal P}(\vec j)=1/(4\pi)$. In terms of  the  harmonic expansion $p(I,\psi)=\sum_{lm} b_{lm}Y_{lm}(I, \psi)$, the coefficients  $b_{lm}$  at $l\lsim 1/\gamma\sim 15$ will be virtually unaffected  by our approximation  (i.e. ignorable at the lower degrees such as $l\lsim  4$).   Geometrically,  Eq. (\ref{tr2}) can be regarded as a $90^\circ$ rotation of the polar direction (from $\theta=0$ to $I=0$), thus keeping the degrees $l$ at the correspondence of  their expansion coefficients (see e.g. \cite{Sakurai:2011zz}). Even if we start from  an axisymmetric model ${\cal P}_A(\cos \theta)$   as in Eq. (\ref{pdf0}),  the transformed one $p(I,\psi)$ can depend on the  polarization (azimuthal) angle $\psi$.  

\section{GW Observation }

\subsection{Waveform Model}
We now focus on a nearly monochromatic GW from a circular CWDB with an orbital frequency $f/2$.    We keep the essential aspects for our study, dropping irrelevant details.  

In  the lowest quadrupole approximation, the gravitational waveform at a given  position is expressed as  
\beqa
{\vect h}(t,{\vec n},I,\psi)&=& A_+(I)\cos(2\pi f t+\alpha){\bf e_+}({\vec n},\psi)\nonumber\\
& & +A_\times (I)\sin (2\pi f t+\alpha){\bf e_\times}({\vec n},\psi) \label{dg1}
\eeqa
with the phase constant $\alpha$ and the two amplitudes 
$A_+(I)\propto (1+\cos^2I)$ and $A_\times(I) \propto  2\cos I$ (see e.g. \cite{pw}).  The transverse-traceless tensors ${\vect e}_{+,\times}({\vec n},\psi)$ are given by   
\beq
{\vect e}_+({\vec n},\psi)={\vec e_k}\otimes {\vec e_k}-{\vec e_j}\otimes {\vec e_j},~{\vect e}_\times({\vec n},\psi)={\vec e_k}\otimes {\vec e_j}+{\vec e_j}\otimes {\vec e_k}  \label{be2}
\eeq
with the  vectors $\vec e_k$ and $\vec e_j$  defined  in Eq. (\ref{be1}).

From Eqs. (\ref{be1}) and (\ref{be2}), we readily obtain the identities
${\bf e_{+,\times}}({\vec n},\psi+\pi)={\bf e_{+,\times}}({\vec n},\psi)$ 
and resultantly 
\beq
{\vect h}(t,{\vec n},I,\psi+\pi)={\vect h}(t,{\vec n},I,\psi). \label{dg0}
\eeq
This degeneracy between $\psi$ and $\psi+\pi$ is fundamental,  originating from  the spin-2 nature of the gravitational radiation.  

From Eqs. (\ref{be1}) and (\ref{be2}), we can also confirm the identities
${\bf e_{+,\times}}({\vec n},\psi+\pi/2)=-{\bf e_{+,\times}}({\vec n},\psi)$ and thus
\beq
{\vect h}(t,{\vec n},I,\psi+\pi/2)=-{\vect h}(t,{\vec n},I,\psi) \label{dg2}
\eeq
(see e.g. \cite{Cornish:2003vj,Roulet:2022kot}).
Correspondingly, the $\pi/2$- rotation of the polarization angle $\psi$ can be effectively absorbed by the phase shift $\alpha+\pi$ in Eq. (\ref{dg1}).  Therefore,  we observationally  have the degeneracy between $\psi$ and $\psi+\pi/2$ at the lowest Newtonian order.  By observing the higher post-Newtonian  waveforms at frequencies $f/2$ and $3f/2$ and measuring their phases relative to that of the Newtonian one (\ref{dg1}), we can, in principle, distinguish the two states at $\psi$ and $\psi+\pi/2$ (see e.g. Eq. (11.295b) in \cite{pw}).   Unfortunately, compared with the Newtonian waveform (\ref{dg1}), the higher ones are suppressed by the post-Newtonian parameter  $\beta^{1/2}=O(c/v)$.  For our CWDBs, we have 
\beq
\beta=  \lmk  \frac{\pi GM_tf}{c^3}\rmk ^{2/3}\sim 10^{-5} \lmk  \frac{f}{{\rm 5mHz}}\rmk^{2/3} \lmk \frac{M_t}{1M_\odot}\rmk^{2/3}
\eeq
where $M_t$ is  the total mass of the binary.   Unlike binary black hole mergers observed by ground-based detectors (see e.g. \cite{LIGOScientific:2020stg}),  the small post-Newtonian waveforms of the CWDBs are easily masked by the measurement noises, and we cannot practically solve  the degeneracy between $\psi$ and $\psi+\pi/2$.
   From Eqs. (\ref{dg0}) and (\ref{dg2}), we have a similar degeneracy between  $\psi$ and $\psi+3\pi/2$.

In summary, GW observation is a geometrical measurement  and intrinsically has a good affinity for studying  the configurations of the Galactic  CWDBs.  However, because of the symmetry of the system,  we  have the fourfold degeneracy between the angles $\psi,\psi+\pi/2,\psi+\pi$ and $\psi+3\pi/2$.  

\subsection{Fourfolded  Distribution Function}
As discussed in the previous subsection, our observable is not the full distribution function $p(I,\psi)$ but the folded one 
\beq
{\bar p}(I,\psi_d)\equiv  \sum_{k=0}^3 p(I,\psi_d+k\pi/2) \label{fold}
\eeq
defined in the parameter ranges
$0\le I\le \pi$ and     $0\le \psi_d\le \pi/2$.  
Interestingly,  unless the order $m$ is a multiple of 4 (e.g. $m=0,\pm4,\pm8,\cdots$), the folding operation erases its $\psi_d$-dependent pattern 
\beq
\sum_{k=0}^3 Y_{lm}(I,\psi_d+k\pi/2)\propto \sum_{k=0}^3 \exp(im k\pi/2) =0. \label{f4}
\eeq
  Correspondingly,  the azimuthal patterns of  the function $p(I,\psi_d)$ come only from $l\ge 4$.  
  
  \add{As commented in Sec. II,   an $l$-th order pattern $Y_{l0}(\theta)$ in the original function Eq. (\ref{pdf0}) generates the term proportional to $L_l(\sin I \cos\psi)$ in the transformed one  $p(I,\psi)$ in Eq. (\ref{tr2}).  
  From Eq. (\ref{eq2}),  we have 
  \beqa
  L_l[\sin I \cos\psi]&+&L_l[\sin I \cos(\psi+\pi)]=0\nonumber \\
  L_l[\sin I \cos(\psi+\pi/2)]&+&L_l[\sin I \cos(\psi+3\pi/2)]=0\nonumber
    \eeqa
    for odd numbers $l$. 
  Thus, the fourfolding operation (\ref{fold}) cancels out all the contributions  from the odd order patterns in Eq. (\ref{pdf0}). 
}

Using Eqs. (\ref{tr2}) and (\ref{fold}) and keeping the even orders $l\le 4$, we obtain 
\beqa
{\bar p}(I,\psi_d)&\propto& Y_{00}-\frac{a_{20}}2 Y_{20}(I)
 +\frac{3a_{40}}8  Y_{40}(I) \label{ffn}\\
& & 
 +\frac{a_{40}}8 \lmk  \frac{35}2\rmk^{1/2} \lkk Y_{44}(I,\psi_d)+Y_{4-4}(I,\psi_d) \rkk \nonumber, 
\eeqa
which is 
symmetric at $I=\pi/2$.  We should point out that the full function $p(I,\psi)$ (generated from Eq. (\ref{pdf0}) by Eq. (\ref{tr2})) has the components $Y_{2\pm2}(I,\psi)$, but they disappear at the folding operation as in Eq. (\ref{f4}) for $m=\pm2$.    

In Fig 2, we plot the folded distribution function (\ref{ffn}) in the coordinate
\beq
(x,y)=2\sin[ I/2] (\cos\psi_d,\sin\psi_d)
\eeq
identical to the Lambert azimuthal equal-area projection centered on the face-on direction $I=0$.  Given the symmetry at $I=\pi/2$, we  present only the range $0\le I\le \pi/2$.

\begin{figure}
 \includegraphics[width=0.99\linewidth]{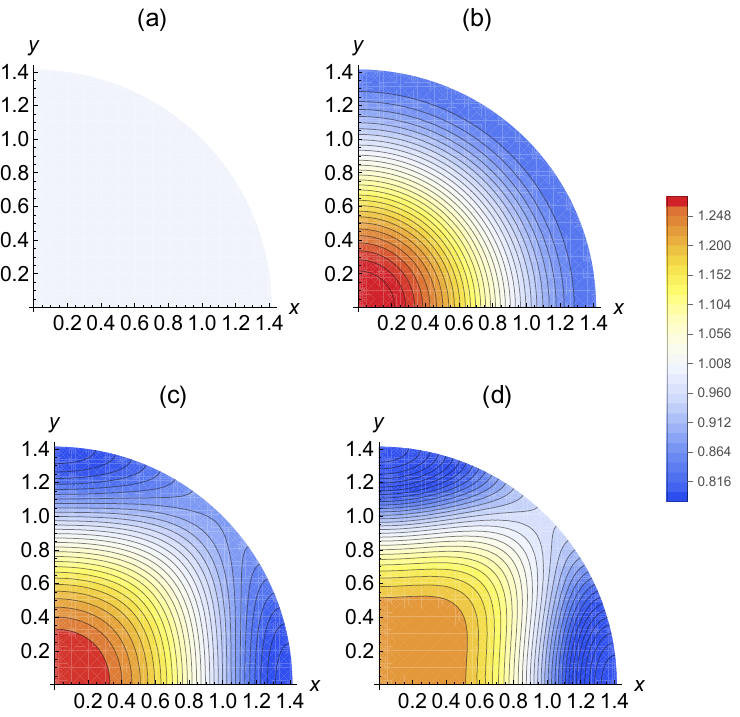} 
 \caption{ The fourfolded orientation  distributions  $\bar p(I,\psi_d)$ (normalized by the factor $\pi^{-1}$).  We apply the area preserving projection $(x,y)=2\sin (I/2)(\cos\psi_d,\sin\psi_d)$. The original distribution functions are axisymmetric model ${\cal P}_A(\cos\theta)$ in Eq. (\ref{pdf0})  characterized by the two anisotropy parameters $a_{20}$ and $a_{40}$.   Panel (a) for the isotropic model  with $(a_{20},a_{40})=(0,0)$,  (b)  with $ (-0.3,0)$,  (c) with $(-0.3,-0.04)$ and (d) with $(-0.3,-0.1)$. 
 }  \label{fig:3}
\end{figure}

For a hierarchical case $|a_{20}|\gg |a_{40}|$,  the azimuthal dependence is weak, as expected from Eq. (\ref{ffn}).  For $a_{20}<0$ (corresponding to the equatorial enhancement in Eq. (\ref{pdf0})),  we have higher probability around the face-on configuration ($I=0$ and $\pi$).

From the folded distribution function $
{\bar p}(I,\psi_d)$, we can easily evaluate the polarization degree of the associated GW background.  The Stokes parameters $(I_s,Q_s,U_s,V_s)$ are its conventional measures \cite{rp,Romano:2016dpx}.   Using the notation $\lnk \cdots \rnk_{\psi_dI}\equiv \int_0^\pi dI \int_0^{\pi/2}d\psi_d \sin I  ~{\bar p}(I,\psi_d)[\cdots]$ for the angular averagings,  we obtain the expressions such as 
\beq
\frac{Q_s+iU_s }{I_s}=\frac{\lnk e^{-4i\psi_d}[ (1+\cos^2I)^2-4\cos^2I] \rnk_{\psi_dI}}{\lnk [(1+\cos^2I)^2+4\cos^2I] \rnk_{\psi_dI}}
\eeq
(defined for the axes $\vec e_r$ and $\vec e_q$).
For the concrete profile (\ref{ffn}), we have
$Q_s/I_s={35a_{40}}/({336-48\sqrt{5}a_{20}+3a_{40}})$ and $U_s/I_s=V_s/I_s=0$.

\section{Probing  anisotropies}

We now discuss how to probe the anisotropies of the original function ${\cal P}(\vec j)$, by analyzing  the observable function ${\bar p}(I,\psi_d)$, which is  sampled by a finite number of CWDBs.   For a probability distribution function defined on a sphere, following e.g. \cite{pro}, we can deal  with the discrete sampling effects on the spherical harmonic expansion.  Here, paying attention to the roughly concentric profiles in Fig. 2,  we rather examine a simple dualistic approach.

We divide our binary sample (in total $N$) into the following two subsets: (i) the low-inclination group with $|\cos I|\ge 1/2$ and (ii) the high-inclination group with $|\cos I|<1/2$.  We put their numbers by $N_{L}$ and $N_H$ ($N_L+N_H=N$) and define the asymmetric ratio by  
\beq
{\cal A}\equiv \frac{N_L-N_H}{N}.
\eeq

For the isotropic (random) profile ${\cal P}(\vec j)= \rm const$, we have the vanishing mean $\lla {\cal A}\rra=0$ and the shot noise $\Delta {\cal A}=N^{-1/2}$ for the very basic binomial distribution.  Below, we conservatively  take the reference number $N\sim 2000$, considering the expected fraction of bulge CWDBs \cite{Ruiter:2007xx}.

For an anisotropic profile ${\cal P}(\vec j)$, we can estimate the mean fractions such as 
\beq
\frac{\lla N_H\rra }{N}=\int_{\pi/3}^{2\pi/3} dI \int_0^{\pi/2}d\psi_d \sin I ~{\bar p}(I,\psi_d)
\eeq
and obtain
\beq
\lla  {\cal A}\rra=-\frac{192 \sqrt{5} a_{20}+135 a_{40}}{1024}\label{rat2}
\eeq
for the function (\ref{ffn}).
 Therefore, if we have the condition 
$
|\lla {\cal A}\rra| > \Delta {\cal A}
$, we can probe the intrinsic anisotropy in ${\cal P}(\vec j)$ (at $1$-$\sigma$ level). Dropping the term $\propto a_{40}$ in Eq. (\ref{rat2}), the inequality can be expressed as 
\beq
|a_{20}|>0.053 \lmk \frac{N}{2000} \rmk^{-1/2}. 
\eeq
This result would serve as a rough guidance for the detection limit of the intrinsic alignment of the bulge CWDBs. 

\add{
We have some comments on the above simple arguments, in view of  the selection effects at actual observational analysis.  
As shown in Eq. (\ref{dg1}), the signal-to-noise ratio $\rho$ depends on the sky position $\vec n$ of the binary and its orientation angles $(I,\psi)$.  For given parameters $({\vec n},\psi)$,  the signal-to-noise ratio $\rho$ generally becomes smallest for an edge-on binary with  $\cos I=0$, as indicated by the expressions $A_{+,\times }(I)$.  How about the dependence on the remaining parameters $({\vec n},\psi)$ for edge-on binaries?  In fact, due to the annual rotation of the detector plane of LISA, the dependence is  effectively averaged out, and  the associated  scatter is largely suppressed.  We can quantitatively examine this, by (i) using the standard framework for  nearly monochromatic binaries with LISA-like detectors \cite{Cutler:1997ta} and (ii) randomly sampling the parameters $({\vec n},\psi)$.  For an  observational period of an integer times 1yr,  the minimum signal-to-noise ratio $\rho$ of the edge-on binaries  is only  $\sim 10\%$ smaller than the rms value of the whole edge-on sample (corresponding to the large number limit of detectors in \cite{Yagi:2011wg}). In any case, the selection effect can be avoided by using CWDBs in the appropriate region in the $(f,{\dot f})$-space so that the Galactic survey is expected to be complete  (e.g. $f>3$mHz) \cite{Seto:2022iuf}.  Alternatively, when preparing the binary sample, we can introduce dependence on the angular variables (${\vec n}, I, \psi$) to  the threshold of the signal-to-noise ratios.
}

Meanwhile, the estimation errors for the parameter $\cos I$ induce missclassifications of binaries  around the boundary  $|\cos I|=1/2$, and 
we might need  numerical studies to  examine  the potential biases for the asymmetric ratio $\cal A$. 
 The Fisher matrix analysis roughly gives $\Delta\cos I\sim 1/\rho$ (except for $\cos I\sim \pm 1$) for a signal-to-noise ratio $\rho$ (typically at $\sim10$-$100$ for a Galactic CWDB) \cite{Seto:2022iuf}.  Therefore, such confusions are relevant for a small fraction ($\sim 1/\rho$) of the binaries, eventually increasing  the shot noise $\Delta {\cal A}$  only slightly (order of $N^{-1/2}\rho^{-1})$.

\section{Discussion and Summary}
So far, we have mainly discussed the orientations of bulge CWDBs.  Using LISA, from the ampluitude and frequency modulations, we can also measure the directions $\vec n$ of the binaries \cite{Cutler:1997ta}. In addition, we will be  able to estimate the distances to some of the inspiraling CWDBs by analyzing their orbital decay rates  ${\dot f}>0$ \cite{Takahashi:2002ky}.  These pieces of positional information will help us to roughly select the bulge components. However, alignment studies will be intriguing also for disk components,  and we do not need to stick too much with a separation between the two components.  In any case, if the observed anisotropies  ${\bar p}(I,\psi_d)$ are turned out to be  strong, we could additionally explore the spatial correlation of the orientations, by combining the two- or three-dimensional positional information.

Let us briefly summarize our study.  Compact binaries emit GWs,  encoding their orbital motions. GW observation is intrinsically geometrical and enables us to measure the configurations of binary sources. Meanwhile, it was recently reported that,    the orientations of the  planetary nebulae associated with short-period bulge binaries are preferentially aligned to the Galactic plane, possibly reflecting the frozen initial conditions of the binaries \cite{Tan}.  
In the near future, LISA will separately detect $\sim 10^4$ Galactic  CWDBs and will become an ideal  tool to examine their  alignments, based only on the first principles of physics. In spite of the fourfold degeneracy of the polarization angles, LISA could enable us to measure the axisymmetric quadrupole pattern  $a_{20}$  as small as  $\sim 0.05$ and could serve as an interesting tool for geometrically delving into the fossil records in our Galaxy.  

\acknowledgments
The author would like to thank M. Iye and K. Tomida for valuable discussions. He also thanks K. Kyutoku and T. Tanaka for useful conversations.

\end{document}